\documentclass[11pt]{article}
\usepackage{amsmath}
\usepackage{graphicx}

\renewcommand{\Re}{{\rm Re}}
\renewcommand{\Im}{{\rm Im}}

\newcommand{\beq}{\begin{equation}}
\newcommand{\eeq}{\end{equation}}

\begin{document}

\title{Trace for the Loewner Equation with Singular Forcing}

\vskip 1 cm

\author{
Leo P. Kadanoff\footnote{LeoP@uchicago.edu}\\
Marko Kleine Berkenbusch\footnote{mkb@uchicago.edu}\\
\\
\parbox{25pc}{\small
\centering
\mbox{The Materials Research Science and Engineering Center}\\
The University of Chicago\\
5640 S. Ellis Avenue\\
Chicago IL 60637, USA\\
}
\vspace{0.5cm}
}
\maketitle
\begin{abstract}
The Loewner equation describes the time development of an analytic map into the upper half of the complex plane in the presence of a ``forcing", a defined singularity moving around the real axis. The applications of this equation use the \emph{trace}, the locus of singularities in the upper half plane. This note discusses the structure of the trace for the case in which the forcing function, $\xi(t)$, is proportional to $(-t)^\beta$ with $\beta$ in the interval $(0, 1/2)$.  In this case, the  trace is  a simple curve, $\gamma(t)$, which touches the real axis twice. It is computed by using matched asymptotic analysis to compute the trajectory of the Loewner evolution in the neighborhood of the singularity, and then assuming a smooth mapping of these trajectories away from the singularity.  Near the $t=0$ singularity, the trace has a shape given by
$$\left(  \Re~ (\gamma(t)-\gamma(0)\right)^{1-\beta}  \approx \left( \frac{\beta}{2 \pi} \Im ~ \gamma(t) \right)
 ^\beta = O   (\xi(t))^{1-\beta}
$$
A numerical calculation of the trace provides support for the asymptotic theory.

\end{abstract}

\newpage
\section{Introduction}
\subsection{Defining Loewner evolution}
O. Schramm \cite{OS} has pioneered a new mathematical approach to the geometry of fractal and/or critical objects in two dimension.  Many different applications have been found to such problems as percolation, random walks, and critical phenomena\footnote{A bibliography can be found in \cite{GK}.}.   The approach has been called Schramm-Loewner evolution (SLE) because it is based upon the earlier work of C. Loewner \cite{LE} on the relation between two-dimensional geometry and differential equations in the complex plane.  This approaches is based upon the relatively simple equation
\begin{subequations}
\begin{equation}
\frac{dg_t}{dt} = \frac{2}{g_t-\xi(t)}
\label{LE}
\end{equation}
containing the forcing function, $\xi(t)$, a real function of a real variable.  Equation \ref{LE}
is solved  with the initial condition
\begin{equation}
g_{t_0} =z
\end{equation}
\end{subequations}
where z is a complex variable that lies in the upper half of the
complex plane, thereby
generating the function
$g_t(z)$. For each value of $t>t_0$, this function can be thought of as a
mapping of the form
$w=g_t(z)$ which takes some connected subset of the upper-half $z$-plane,
$\mathcal{R}$, into the entire region above the real axis of the $w$-plane, $\mathcal { U}$.
It also maps the curve bounding $\mathcal{R}$, called $\Gamma$, onto the
real line of the
$w$-plane.

The Loewner equation continually generates new singularities of $g_t(z)$.
At each time, $s$, a new singularity is generated at the point, $z=\gamma(s)$, such that
\beq
g_s(\gamma(s))=\xi(s).
\label{SING}
\eeq
As time goes        on, points in the upper half-plane are continually added to\footnote{If $\gamma$ is a simple curve that intersects the real axis once, $\Gamma$ is the union of that curve and the real axis.}.  $\gamma$ and $\Gamma$.
The set $\Gamma$ is a geometrical object which has evoked considerable interest, particularly in the case in which the forcing, $\xi(t)$ is a Brownian process \cite{GK}.

\begin{figure}[t]
\includegraphics[scale=1.0]{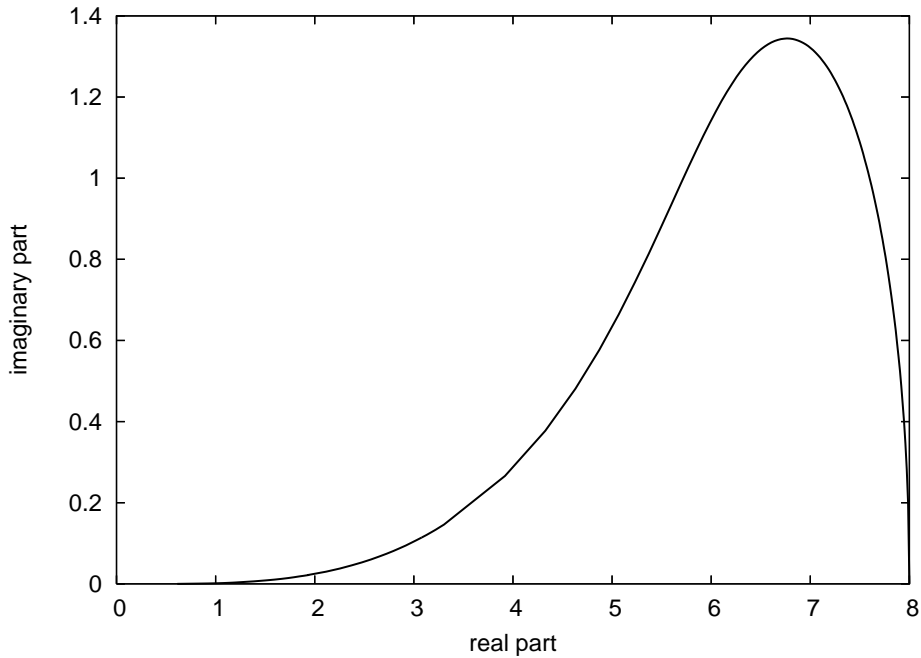}
%\vskip 1.5 in
%\includegraphics[scale=.35]{wplane.pdf}
%\vskip 1 in
\caption{This figure shows the trace  for the case of power law forcing for $\beta=1/4, \kappa=8$ and $t_0=-1,0$. The right hand side of the trace curve is a nonsingular orthogonal intersection with the real axis.   On the left the singularity in the forcing has generated a singular osculation with the axis. }
\label{BetaFig}
\end{figure}

The function $\gamma(t)$ has some very remarkable geometrical properties, all derived from the nature of the singularities in $\xi(t)$.  For the purposes of this paper, the most important such property is the non-self-touching behavior of $\Gamma$.  If $\xi(t)$ has all its singularities weaker than $(t_c-t)^{1/2}$ (for example if it has a derivative everywhere) then $\gamma(t)$ is a simple curve which never touches itself.  Conversely if $\xi(t)$ has a singularity stronger than the one half power, e.g.  $(t_c-t)^{\beta}$ with $0<\beta<1/2$ then $\Gamma$ will intersect or touch itself \cite{MR}.   These touchings (or non-touchings) are crucial to the descriptions of the critical processes which are the main applications of SLE.  One example of such a touching trace is shown in figure \ref{BetaFig}.  It rises on the left as a square root of time, but is there non-singular in its shape.  On the right is a singular osculation with the real axis.  This paper is devoted to finding the form of this singularity.

\subsection{Orbits}
To calculate the trace we need to find the
 \emph{orbits} or \emph{trajectories} produced by equation \ref{LE}. Thus we look at  $g_t(z)$ while varying $t$ with $z$ held fixed.  One family of such orbits runs from the trace to the real axis.  It is formed by setting $z=\gamma(s)$ where $s$ is a real variable with the meaning of a time.  We demand $s>t_0$. Then the orbits we consider  are the curves traced out by $g_t(\gamma(s))$ when $t$ is varied through the interval $[t_0,s]$.  At the one end, the curve hits  points on the trace, since
\beq
g_{t_0}(\gamma(s))= \gamma(s),
\label{FIND-gamma}
\eeq
On the other end, the curve  intersects the real axis (see equation \ref{SING}). The intersection is orthogonal if $\xi(t)$ is non-singular at  $t=s$.
To find the trace \cite{MR} use equation \ref{SING} to provide  initial conditions for all  the trajectories $g_t(\gamma(s))$ and  run the evolution, equation \ref{LE} , backward from time $s$ to time $t_0$.  For each $s$, the solution at that time determines $ \gamma(s)$  via equation \ref{FIND-gamma}.  One family of such orbits is depicted in figure \ref{LINFig}.

\begin{figure}[ht]
\centering
\hskip 0.5 in
\includegraphics[scale=.450]{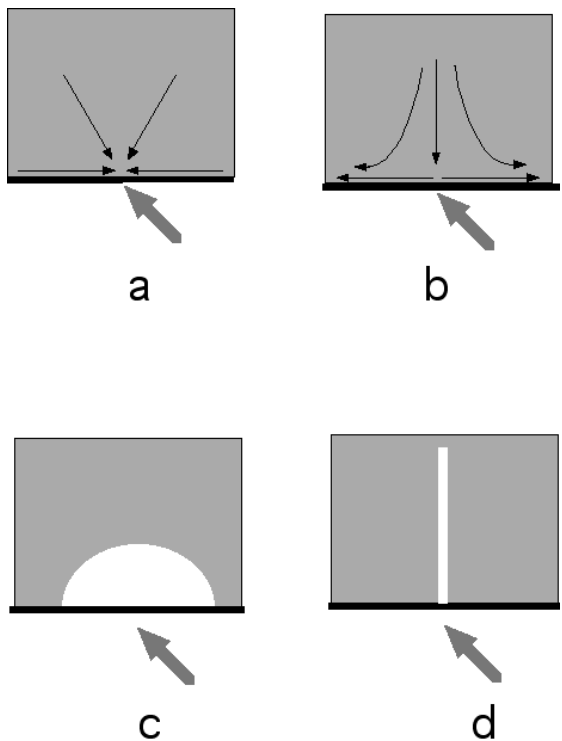}
%\vskip 1.5 in
%\includegraphics[scale=.35]{wplane.pdf}
%\vskip 1 in
\caption{These four panels show a ``fluid" in grey.  The grey arrow points to a singularity or leak on the boundary.  In panel a, the flow is the one produced by a point sink; in b, the flow pattern is the one used in this paper (see equation \ref{LE} ) in which there is downflow in the vertical and outflow in the horizontal direction. The white regions in the next two panels show the regions containing the fluid which will leave the system through the leak.}  \label{FLOW}
\end{figure}

This kind of analysis is familiar from the theory of the motion of ``passive scalars" \cite{PS}. Take  equation \ref{LE} to describe the motion of a point in the $w-$plane driven by a velocity field $2/(w-\xi)$.  At first sight,  this appears to be just the motion engendered by a point source or sink at $w=\xi$.    However that interpretation is wrong.  A point sink at $\xi$ would generate the field $-2/(\bar{w}-\xi)$ where the bar stands for complex conjugation.  Figure \ref{FLOW} shows the difference between the two flow patterns.  The conjugated field has an outflow per unit time of fluid occupying one unit of area. Thus after a time, an entire area of the fluid has escaped.  The actual flow of equation \ref{LE} directly allows a much more restricted escape.  If the forcing is sufficiently smooth, everywhere smoother than Hoelder index $1/2$, the escaped points form a simple curve which intersects the real axis at $\xi(t_0)$.

\subsection{Outline of paper}
This note describes the singular properties of equation \ref{LE} when the forcing is singular and has the form
\begin{equation}
\xi(t) =2 \sqrt{\kappa} ( -t)^\beta   \quad
\end{equation}
for the case in which $\beta$ is in the interval $(0, 1/2)$ and $t_0$ is smaller than zero.  To understand the various different  $t=0$ singularities generated by the Loewner evolution, we concentrate upon the trace, $\gamma(t)$. All other singularities are derived from the ones in $\gamma(t)$.

    Both trajectories and trace  can be studied analytically \cite{KNK} for $\beta=0, 1/2,$ and $1$.  In this paper we study  singularities for $\beta$ in the interval $(0,1/2)$ by performing an asymptotic analysis of the  properties of trajectories, $g_t(z)$,  for $t$ and $z$ in the neighborhood of the singularity.

The next section of this note is devoted to describing the exact solutions. and the orbits which arise from them.  Results for $\beta=1$ are particularly needed because they are used in the subsequent small-$\beta$ analysis.   Section \ref{SingTraj} turns to the situation for $0<\beta<1/2$, which is studied via matched asymptotic analysis \cite{HINCH}.  This part looks at the behavior of the critical trajectory, the one which passes through $z=0$, and at the linear stability analysis about that trajectory.  The critical trajectory has the unique property that it intersects the real axis twice.  Section \ref{NearCrit} is devoted to describing the small $z-$ trajectories near their intersection with the real axis.  The final section connects the behavior of these trajectories to the linear analysis of section 3 and thereby finds the actual singularity in the trace.    The trace is found asymptotically when $-t_0$ is a positive number much smaller than one. Away from the singularity at time zero,  the time development is described by smooth mappings, and consequently we argue that leading order shape of the trace is universal and applies, for example, to larger values of $-t_0$.

\section{Exact Analysis} \label{exact}
\subsection{The case $\beta =0$.}
Pick $\xi(t)=0$.  Then the solution for $g$ is simply
\beq
g_t(z) = \sqrt{z^2 +4(t-t_0)}
\eeq
and the trace lies on the imaginary axis, having $\gamma(s)=2i \sqrt{s-t_0}$.   As a result the orbit leading up to the singularity is
\begin{equation}
g_t(\gamma(s)) = 2i \sqrt{s-t}      \quad  \mbox{for}  \quad   t_0 < s  < t
\end{equation}
In this case,  for all values of $s$ the singular trajectories  $g_t(\gamma(s)) $ follow the imaginary axis.

\subsection{The case $ \beta=1$.}
Next, consider the case of a linear forcing:
\beq \xi(t)=t_0-t
\label{LIN-def}
 \eeq
 which was fully solved in reference \cite{KNK}.        The linear solution\footnote{Here we use a convention for the sign of $t$ in equation \ref{LIN-def} different from the one used in ref \cite{KNK}.  This results in sign changes in equation \ref{FDEF} compared with the corresponding equation in that paper.}  depends upon the function F which is defined by
 \begin{equation}
F(z)=-z+2 \ln(1+z/2)
 \label{FDEF}
 \end{equation}
This paper will make use of the functional inverse of $F$, i.e. the function $F^{-1}(x)$, with complex $x$ in the upper half plane,  which obeys $F^{-1}(F(z))=z$. For $x$ with small magnitudes,
\begin{subequations}
\beq
F^{-1}(x) =  2 i \sqrt{x} -2x/3 +\frac{13 i }{18} x^{3/2} +\cdots
\label{smallF}
\eeq
The branch is picked so that the square root is positive when $x$ approaches the  real axis.
If the real part of $x$ is large  then the corresponding branch gives
\beq
F^{-1}(x) = -x +2[ \ln (x) + \pi i   -\ln 2] [1-2/x] -4/x + \cdots
\label{largeF}
\eeq
\end{subequations}

\begin{figure}[t]
%\includegraphics[scale=.55]{linear.eps}
%\vskip -1 in
\includegraphics[scale=.52,angle=270]{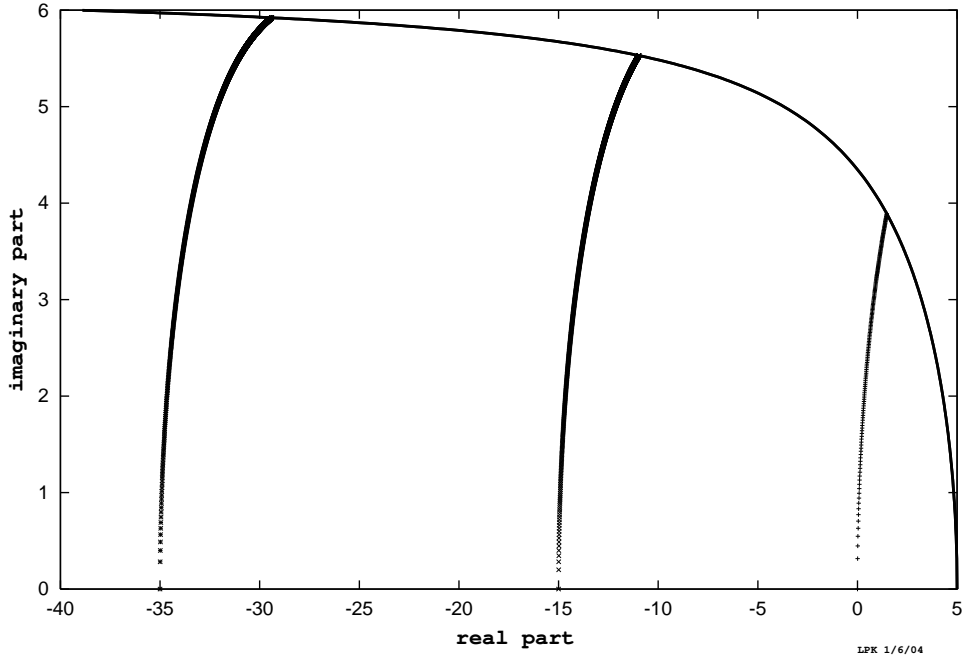}
%\vskip 1 in
\caption{This figure includes the trace and three trajectories for the case of linear forcing. Specifically, we take $\xi(t)= (5-t)$ and $t_0=0$.  The solid line at the top is the trace. The symbols  define trajectories with. counting from the left,  $s=5, 20, $ and $40$.  }
\label{LINFig}
\end{figure}

The solution for $g$ is given in terms of this known function, $F$, by
\beq
F(g_t(z)-\xi(t)) =F(z) +\xi(t)
\label{OLD-LIN}
\eeq
The trace is given by setting $z=\gamma(t)$ with $\gamma$ being picked  so that $g_t(\gamma(t)) =\xi(t)$. In this situation, this identification makes the left hand side of equation \ref{OLD-LIN}  equal to $F(0)= 0 $, and consequently the trace obeys
$$
F(\gamma(s))= -\xi(s)=s-t_0
$$
with the final consequence
\beq
\gamma(s)= F^{-1}(-\xi(s)).
\label{OLD-TRACE}
\eeq
On the other hand, the critical trajectory  is obtained by setting $z=\gamma(s)$ in equation \ref{OLD-LIN}, which then reads
$$
F(g_t(\gamma(s))-\xi(t)) =F(\gamma(s)) +\xi(t)=s-t
$$
This equation may then be solved to find the orbit in terms of the trace.
\beq
g_t(\gamma(s)) = \xi(t)+F^{-1}(s-t)
\label{OLD-ORBIT}
\eeq
Figure \ref{LINFig} shows the behavior of the trace and some of the trajectories.  The trajectories connect the trace-points with the  corresponding singular points on the real axis. In fact, of course, the trace is calculated as the locus of end-points of the trajectories.   At the intersections of the orbits with the real axis, the curves meet the axis orthogonally. (Also see equation \ref{LINSmall} below.) This behavior is required whenever $\xi(t)$ is non-singular at the intersection.  The points plotted on  the orbits are all equally spaced in $t$.  The large spacings on the figure near the intersections reflect the square-root singularities at these points.       The figure hints that even for large $s$ values the trajectories remain bounded below $y=2 \pi$, and only grow slowly in their overall length. The exact solution (see equation \ref{Larget-traj} below) shows that this growth is, in fact, logarithmic in $s$.

The results in figure \ref{LINFig} can be seen more precisely by calculating the asymptotic properties of trace and orbit from the exact solution.   For $s$ just bigger than $t_0$,  equation \ref{OLD-TRACE}
 implies that the trace has the value
 \begin{subequations}
 \beq
\gamma(s)=2 i\sqrt{s-t_0}  -2(s-t_0)/3  +\cdots
\eeq
while from equation \ref{OLD-ORBIT}  as $t$ approaches $s$ from below (and $s>t_0$), i.e. just before the trace hits the singularity,  the orbit obeys
\beq
g_t(\gamma(s)) = t_0-t+2 i\sqrt{s-t}  -2(s-t)/3 +\cdots
\label{LINSmall}
\eeq
\end{subequations}
In the opposite limit, i.e. as time, $s$, goes to infinity, the trace obeys
 \begin{subequations}
\beq
\gamma(s)  =-s+t_0 +2 \ln [(s-t_0)/2] + 2 \pi i      +O (   \frac{\ln (s-t_0)}{s-t_0}  )
\label{Larget-trace}
\eeq
while, for large $s-t$  the orbit approaches the limit
\beq
g_t(\gamma(s) ) = t_0 - s+ 2[\ln [(s-t)/2] +2 \pi i][1-2/(t-s)]   +\cdots
\label{Larget-traj}
\eeq
\end{subequations}
Equation \ref{Larget-trace} shows as $s$ becomes very large, the trace moves with almost  the same velocity as the point of forcing, $\xi(s)$, but it always lags slightly behind it. Asymptotically, the separation between trace and $\xi$ goes to a constant, $2 \pi $, in the imaginary direction and  grows only logarithmically in the real direction.       Correspondingly, equation \ref{Larget-traj} shows that the orbit covers this separation in the time between $t_0$ and $s$.

\subsection{The case $\beta=1/2.$}
So far we have seen no intersections  or touching  of trajectories and the real axis, except for the trivial take-off at $t_0$.   With a forcing of the form $2 \sqrt{\kappa(1-t)}$, and $\kappa>4$, there is an exact solution \cite{GK} which exhibits a non-trivial intersection with the real axis. The intersection is approached by a trace in the form of straight lines at an angle to the real axis. The angle is expressed  in terms of the parameters \cite{GK}, $y_\pm$
\beq
y_\pm =  \sqrt{\kappa }\pm \sqrt{4-\kappa}
\label{slope}
\eeq
The angle is given as $\phi$ where
\beq  \phi = \pi y_- / y_+ .
\label{phi}
\eeq

\begin{figure}[t]
%\includegraphics[scale=.55]{linear.eps}
%\vskip -1 in
\includegraphics[scale=.4, angle=270]{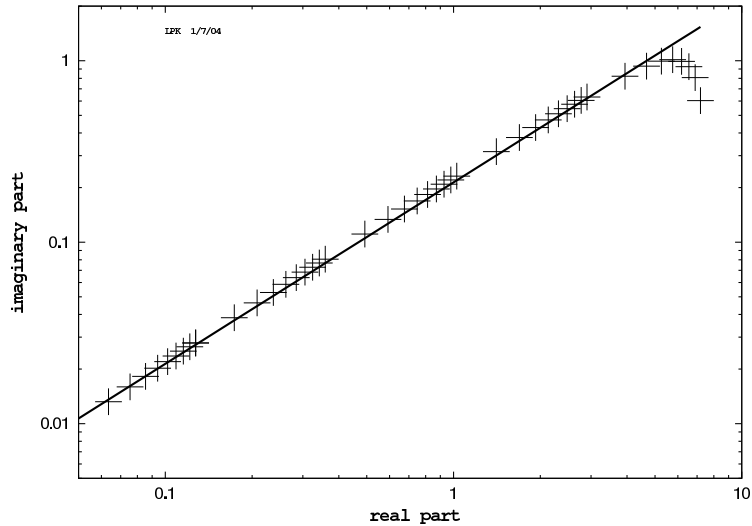}
%\vskip 1 in
\caption{This figure compares the trace for a square root singularity and $\kappa=16$ with its small-$s$ fit.  The numerical data is the points; the line is the fit with slope and intercept chosen to be the right theoretical values.  The fit looks and is excellent. }
\label{HalfFig}
\end{figure}

This trace and its comparison with the exact solution are shown in figure \ref{HalfFig} for the case in which $\kappa=16$.
At $s=0$, the value of $\gamma(s) $
is equal to $\xi(s)$  and is hence $8$.  This value is the point on the far right of the trace curve.  From this point, as $s$ increases, the imaginary part of the trace grows
 and then falls away. Over the entire range of the plot, the real part of the trace is roughly proportional to $\xi(s)$.  Finally as $s$ approaches the singularity at $1$ the imaginary part of the trace decreases and goes to zero proportionally to $\xi(s)$, In this limit,  the real part also varies as $\xi$. Thus as $s$ approaches $1$ the trace has the behavior
\beq
\gamma(s)- y_-=C    e^{i\phi } \xi(s)
\label{TRACE1/2-sing}.
 \eeq
 where $\phi$ is given by equation  \ref{phi}, $C$ is real and positive,
 and $y_-$ is the value of the trace at zero found from the exact solution.

% \begin{figure}[t]
%%\includegraphics[scale=.55]{linear.eps}
%%\vskip -1 in
%\includegraphics[scale=.52,angle=270]{fig3LIN.eps}
%%\vskip 1 in
%\caption{This figure includes the trace and four trajectories for the case of forcing with $\xi(t)=8 (1-t)^{1/2}$ and $t_0=0$.  The solid line at the top is the trace. The symbols  define trajectories with, counting from the right,  $s$=0.2, 0.4, 0.6,  0.8, 0.9, and 0.99. }
%\label{HalfLINFig}
%\end{figure}

Figure \ref{HalfFig} was calculated numerically, rather than from using the exact solution.   Since there is a singularity at $s=1$,  the calculation could not be carried all the way down to $s=1$.  However the calculation did continue quite close to the exact intersection point at $\gamma(1) = y_-  \approx 0.53$, specifically to a time within $10^{-5}$ of the critical time. The calculation is rather straightforward and is described in the appendix.  Since the exact solution is available, a comparison to the exact solution can be made, giving a relative accuracy of 3 parts in $10^{8}$ at this closest trace point to the singularity, and better accuracy further away.   This curve also shows that the numerically calculated result is fit over a quite reasonable range by the near-singular asymptotics.

 \section{The Critical Trajectory} \label{SingTraj}
 Here we discuss the singularities in Loewner evolution for the case in which the forcing has a singularity with index $\beta$ at the time $t=0$.  As we shall see, in the studied interval, $0<\beta<1/2$,   for values of $t$  near zero the critical trajectory has $g_t$ much smaller than $\xi(t)$.  Hence this trajectory and nearby ones can be studied by expanding in the $g$ on the right hand side of equation \ref{LE}.   This expansion produces an analytically tractable problem which is studied in this section.

 \subsection{Behavior of singular trajectory}
We solve equation \ref{LE},  going backward in time from the ``initial" condition
\begin{equation}
g_0(\gamma(0))=0
\end{equation}
We notice that $\xi$ is changing quite rapidly in the neighborhood of $t=0$ so that we might expect that $g_t(\gamma(0))$ is much smaller than $\xi(t)$ in this neighborhood.  Thus we might try an expansion of  equation \ref{LE} of the form
\begin{equation}
\frac{d}{dt}g_t(\gamma(0))= -2/\xi -2g_t(\gamma(0))/\xi^2  -2(g_t(\gamma(0)))^2/\xi^3 + \cdots
\end{equation}
This equation can be solved iteratively with the lowest order solution being
$$
g_t(\gamma(0))= \frac {2 ~ (-t)^{1-\beta}}{1-\beta} +\cdots
$$
and the next order
\begin{equation}
g_t(\gamma(0))= \frac{2 ~ (-t)^{1-\beta}}{1-\beta} + \frac {4 ~ (-t)^{2-3\beta}}{(1-\beta)(2-3\beta)} +\cdots
\label{CRIT}
\end{equation}
This trajectory is the one which causes the trace to hit the real axis.   It has this effect because it remains real. This reality is the expected behavior for a trajectory which describes a point at which the trace touches or intersects itself.   On the other hand, all the other trajectories pop off the real axis as soon as $-t$ increases beyond its initial value,  $-s$.

Convergence of this series requires $\beta$ to be less than one half so that exponents of the form $n-(2n+1)\beta$ do not change sign.

Note also that our initial assumption that $g_t  \ll \xi(t) $ is quite true for this trajectory since
$$(-t)^{1-\beta}  \ll (-t) ^\beta $$
for small values of $-t$ and $\beta$ in the interval $(0,1/2)$.

Thus equation \ref{CRIT} is accurate when $-t \ll1$.  If it is also true that $-t_0 \ll 1$. we can find find the $g$-value at the initial time $t_0<0$ and hence find the trace at $s=0$ as
\beq
\gamma(0)= g_{t_0}(\gamma(0))= \frac{2 ~ (-t_0)^{1-\beta}}{1-\beta} + \frac {4 ~ (-t_0)^{2-3\beta}}{(1-\beta)(2-3\beta)} +\cdots
\label{TR-SOLN}
\eeq
This then tell us the position of the end of the trace, but nothing about the shape of the curve $\gamma(s)$.  Further, we do not know the trace even at $s=0$ when $-t_0$ is not small.

\subsection{Near-critical behavior: Linear deviations from critical trajectory } \label{NearCrit}
To see some of the properties of a trajectory near the critical one, take $s$ to be small and write
\begin{equation}
g_t (\gamma(s) )= g_t (\gamma(0)) + \Delta_t .
\end{equation}
Then  the deviation $\Delta_t  $ must in some sense be small.  In a linear analysis, it obeys
\begin{equation}
\frac{d}{dt} \Delta_t  = - 2 \Delta_t /[g_t (\gamma(0) )-\xi(t)]^2
\label{LEDel}
\end{equation}
with the initial condition that
\beq
 \Delta_{s}  = g_s (\gamma(s) )- g_s (\gamma(0)) \approx (-s)^\beta - \frac{ 2(-s)^{1-\beta}}{1-\beta}
\label{BCLEDel}
\eeq

However, we cannot use this initial condition here.  Because the first term dominates the right hand side of equation \ref{BCLEDel},  $ \Delta_t $ is not a small perturbation near $t=s$ and we cannot employ   equation  \ref{LEDel} in the region.  Thus we cannot make direct use of the initial data of equation \ref{BCLEDel} in our analysis of equation \ref{LEDel}.

On the other hand, for larger values of $-t$, i.e. specifically for some negative $t$ in the range. $-s \ll -t \ll 1$ equation \ref{LEDel} does apply.  Under the stated conditions, $g$ is much smaller than $\xi$ so that the equation takes the simple form
 $$\frac{d}{dt} \ln \Delta_t =  - 2 [\xi(t)]^{-2}
$$
which then has the solution
$$
g_t (\gamma(s) )= g_t (\gamma(0)) + \Delta_T \exp[   2 \frac{(-t)^{1-2\beta}} {1-2\beta}
-   2 \frac{(-T)^{1-2\beta}}{1-2\beta}]
$$
where $T$ is some reference time in the interval $-s \ll -T \ll 1$.
Since $T$  has no physical meaning, one can absorb the factor involving $T$ into $\Delta_T$ and get a result of the form
\beq
g_t (\gamma(s) )= g_t (\gamma(0)) + \Delta \exp[   2 \frac{(-t)^{1-2\beta}} {1-2\beta} ]
\label{DEL-SOLN}
\end{equation}

Equation \ref{DEL-SOLN} is accurate only when one can expand in the correction term in $\Delta$.   If $-t_0$ is bigger than or of the order of  one and $-t \ll 1$, the explicit $t$-dependence in the exponent of  Equation \ref{DEL-SOLN} does not matter, and the equation is accurate whenever
\beq
|\Delta|  \ll  |\xi(t)|   \quad \mbox{and} \quad -t \gg -s
\label{Crit-Lim}
\eeq

     To complete this analysis and find our trace, we have to extrapolate to  values of $t$ closer to $s$, and then used the boundary condition of equation \ref{BCLEDel}.  This analysis is the job of the next section. The final section brings the various results together.

\section{Near-critical trajectories close to their intersection with the real axis} \label{Linear}
In the previous section, we looked at situations in which $\xi(t) \gg g_t$.   Now we wish to focus upon the orbits' intersections with the real axis.   At intersection, $\xi(t)=g_t$; near intersection they are close to one another.  So an expansion of the right hand side of equation  \ref{LE}  in   $g_t$ will not work.

On the other hand, near enough to the intersection we can expand $\xi(t)$ in a power series around $t=s$.  Because the linear case is exactly solvable, we can expect to get accurate solutions    (albeit with only a limited range of validity) by using this approach.

The trace, $\gamma(s)$, has been defined so that the trajectory $g_t(\gamma(s))$ intersects the real axis at  $t=s$. Take  $s$ to be a small negative number.  In the neighborhood of the intersection the forcing has the value
\begin{equation}
\xi(s-\tau) \approx \xi^L(s-\tau) \equiv \delta+  \tau  \Lambda     \quad \mbox{for  } \quad  0 \leq \tau \ll -s
\label{LIN}
\end{equation}
where the constants $\delta$ and $ \Lambda $ are respectively very small,
\begin{subequations}
\beq
\delta=2 \sqrt{\kappa}(-s)^\beta.
\eeq
 and very large
 \beq
 \Lambda=2 \sqrt{\kappa} \beta(-s)^{\beta-1}
 \eeq
 \end{subequations}

\subsection{Solution of equation with linearized forcing.}
Equation \ref{OLD-LIN} gives the solution when the forcing is just $-t$.
Our case is slightly different for two reasons.  First the forcing is not just $-t$ but a linear function of $t$.  Secondly because the solution derived in this section only applies for small $\tau$, we cannot directly apply a boundary condition in $z$.  To make these changes, we use the invariances in ref \cite{KNK} to see that the solution takes the form:
$$
F\big(\Lambda[g_t(z)-\xi^L(t)]\big) =  c(z) +  \Lambda   \xi^L(t)
$$
where $F$ is given by equation \ref{FDEF}, $c(z)$ is a ``constant" of integration, and $\delta$ and $ \Lambda $ are both independent of $\tau$.
To set the constant of integration for the trajectory, note that at the intersection of the trajectory with the real axis, $\xi=\delta$, while the argument of $F$ vanishes and so does  $F$ itself .
Thus we find
$$
F\big(\Lambda[g_{s-\tau}(\gamma(s))-\delta -\Lambda \tau] \big) =    \Lambda^2   \tau
$$
which then has the solution
\beq
g_{s-\tau}]\left(\gamma(s)\right)=\delta+\Lambda \tau   +\Lambda^{-1}  F^{-1} (    \Lambda^2   \tau)
\label{LIN-SOLN}
\eeq
This result can be stated explicitly in two limits.  From equation \ref{smallF} we find
\begin{subequations}
 \begin{equation}
g_{s-\tau}(\gamma(s))=\delta+\Lambda \tau/3   + 2i   \sqrt{  \tau}   \quad \mbox{for}   \quad 0<\Lambda^2   \tau \ll 1
\label{LIN-small}
\end{equation}
The opposite limit comes from equation \ref{largeF} which then gives us the limiting form
\begin{equation}
g_{s-\tau}(\gamma(s))=\delta+\Lambda^{-1}[ 2 \pi i+  \ln (\Lambda^2 \tau/2)    ]    \quad \mbox{for} \quad  \Lambda^{-2}  \ll  \tau \ll  -s
\label{LIN-large}
\end{equation}
\end{subequations}

Equation \ref{LIN-large} is the main result of this section.  This equation enables us to see that for $\tau$ much larger than $|s|$, to leading order,  $g$ contains
\begin{itemize}
\item  a large real $s-$dependent term ($\delta= 2\sqrt{\kappa}(-s)^\beta$)
\item  a much smaller imaginary  $s-$dependent term, and
\item a real term with a weak $t-$dependence.
\end{itemize}
The last dependence is weak both because the coefficient is small and because a logarithmic dependence is a weak one.

\section{Connections}  \label{CONN}

\subsection{A small difficulty}
This paper has concerned itself with two different expansions, the expansion about the critical trajectory of section \ref{SingTraj} and the linear expansion of $\xi$ in section \ref{Linear}.  In the language of matched asymptotics \cite{HINCH}, the latter is a sort of inner expansion, the former a sort of outer one. So in the usual approach, we determine the overlap region of the two expansions, determine the parameters on one in terms of those on the other side, and declare the problem solved.  The linear expansion for $g_t(\gamma(s))$ is acceptable so long as $\tau=t-s  \ll s$.  Conversely the expansion about the critical trajectory is sensible when
(see equation \ref{Crit-Lim} ) the deviation parameter $\Delta$ from the critical trajectory obeys
$|\Delta |  \ll  \xi(t)$.   The most reasonable guess for $\Delta$, which is a deviation from criticality for $g_t(\gamma(s))$, it to choose it to be the biggest constant term in $g$ as it was determined in the linear expansion, namely $\delta=\xi(s) \sim (-s)^\beta$.  Now that result presents a problem.  Our deviation from criticality argument is correct only if $ \tau >>-s$, exactly the opposite of the range permitted by the expansion argument.  We now conclude that there is no overlap of regions of validity of the two expansions, and consequently we must do better to match up our asymptotics.

\subsection{A better expansion}
This section will make heavy use of the result of equation \ref{LIN-large} which says that, in an appropriate range, $g_t(\gamma(s))$ is almost independent of $t$.  The starting point will be a consequence of equation \ref{LE} that
$$
g_t(\gamma(s))=g_T(\gamma(s))  -  \int_T^t  2~du/[\xi(u)-g_u(\gamma(s))]
$$
Take $T$ to lie in the asymptotic large-$\tau$ range of the linear expansion (see equation \ref{LIN-large}) and substitute the known form of $g$ in this range for $g_T$.  Thus our integral expression for $g$ becomes
\begin{eqnarray}
g_t(\gamma(s))=& \delta+\Lambda^{-1}[ \ln (\Lambda^2 (s-T)/2)  +  2 \pi i ]      -  \int_T^t  2~du/[\xi(u)-g_u(\gamma(s))]   \nonumber  \\  & \quad \mbox{for} \quad  \Lambda^{-2}  \ll s-T \ll  -s
\label{int-eqn}
\end{eqnarray}

To understand the integral equation, notice that the integral has a
logarithmic singularity near its lower cutoff at $T$.  This singularity
arises  because  $g_u$ is to lowest order constant, independent of $u$,
while $\xi(u)$ can be replaced by $\delta(s)$ plus a linear term in
$s-u$.  So long as $t$ remains in this linear regime, we continue to
have a behavior in which the integral grows  as $\log t$ However, as
$t$ gets larger, and hence the range of $u$ gets larger, the
denominator ceases to be a linear function toward the upper limit of
the integral.  In fact, for $u-s$ equal to or of the order of $-s$,
quadratic and higher  terms in the expansion of  $\xi(s+(u-s))$ in
$u-s$ increase the size of the denominator and hence cut off the
integral.  As that happens the integral increases only very slowly with
$t$, and we say that  the logarithmic singularity has been cut off.
In this range of $t$, we can say that $g_t(\gamma(s))$ has become
constant,  and takes on the value

\begin{eqnarray}
g_T(\gamma(s)) &=& \mu(s) +O(1/\Lambda)
\label{g-const} \\
    \mu(s) &=& \delta+\Lambda^{-1}[ \ln (\Lambda^2 (-s) /2) +  2 \pi i ]
\label{MU}
\end{eqnarray}

Now apply equation \ref{int-eqn} to a somewhat higher value of $-T$, i.e. $s-T$ of order $-s$.  We can safely replace the $g$'s by the constants $\mu$, at the cost of a small error.  The result is
\begin{eqnarray}
g_t(\gamma(s))=&\mu(s)    -  \int_T^t  2~du/[\xi(u)-\mu(s)] +O(1/\Lambda)  \nonumber  \\  & \quad \mbox{for} \quad s-T=O( -s)
\label{int-eqn1}
\end{eqnarray}
Expand to first order in $\mu$, do the integral, and find
\beq
g_t(\gamma(s))=\mu(s)    +2 \frac{ (-t )^{1-\beta}} {1-\beta}  +2 \mu(s) \frac{ (-t )^{1-2\beta}} {1-2 \beta}+  O(1/\Lambda)
\label{SOLN}
\eeq

Equation \ref{SOLN} is the basic result of this section and indeed this paper.  It is accurate whenever the denominator $\xi(u) -g_u(\gamma(s))$ always has $g$ neligeble in comparison to $\xi$ through the region in which $u$ is of order $t$.   This in turn will be true whenever $-t$ is very small in comparison to unity.  Since equation \ref{SOLN} demands that the linearized solution be in its large ``$x$" asymptotic range we must have the conditions
\beq
 \Lambda^{-2}        \ll   -t+s \ll  1
\eeq

Equation \ref{SOLN} represents a small deviation for the critical trajectory. As such it should be, and is, identical  in form to our earlier result (equation \ref{LEDel}) for the deviation.  The only change is the the unknown parameter $\Delta$ in the earlier result has been replaced by the known parameter $\mu(s)$.   For all  values of $t$ near the critical time, $-t \ll 1$,  the final term in equation \ref{SOLN} is negligible compared to the first.  The second term is large when
 $$ -t \gg   (-s)^{\beta/(1-\beta)}$$

  \subsection{Evaluations of the trace.}
If $-t_0$ is much smaller than one, equation \ref{SOLN} is accurate and one can get the value of the trace simply by setting $t=t_0$ where $g_t(\gamma(s))=\gamma(s)$. We thus find
$$
\gamma(s)= \mu(s) [1+O((-t _0)^{1-2\beta})]
    + \frac{ (-t_0 )^{1-\beta}} {1-\beta}  +  O(1/\Lambda)
$$

Here the symbol $O$ indicates real correction terms.
Our goal is to find the shape of the trace.  We use this expression to
form $(\gamma(s)-\gamma(0))$, using $\mu(0)=0$, and then take the real
and imaginary part of the result, to find
\beq
\Re (\gamma(s)-\gamma(0))  \approx \xi(s) \approx \left[\frac{\beta}{2 \pi} ~ \Im \gamma(s)\right]^{\beta/(1-\beta)}
\label{RESULT}
\eeq

 This result can be extended to values of $-t_0$ of order unity or larger.  The mapping that extends the trajectories further toward negative $t$ values is conformal and hence differentiable throughout the upper half plane.  As such it modifies any short piece of curve, e.g. the near-critical trace,  by displacing it, magnifying it by the constant factor, $C$, and rotating it.  Because the $s=0$ point on the trace sits on the real axis, the displacement is parallel to the axis and the rotation is nil.  Consequently  for earlier values of the initial time, $t_0$, equation \ref{RESULT} is replaced by
 \beq
\Re (\gamma(s)-\gamma(0))   \approx [\beta ~ \Im \gamma(s)]^{\beta/(1-\beta)} \approx C   \xi(s)
\label{RESULTC}
\eeq

\begin{figure}[ht]
\centering
\includegraphics[scale=0.85]{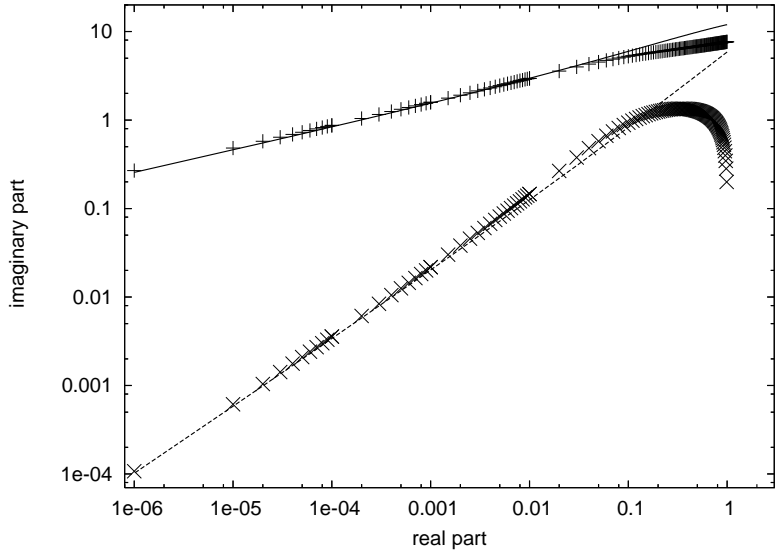}
%\vskip 1.5 in
%\includegraphics[scale=.35]{mul.eps}
%\vskip 1 in
\caption{This figure plots the real and imaginary parts of the trace against the time interval to the singularity.  The upper set of points  is the real part; the lower the imaginary part.  For comparison the curves give  the lowest order asymptotic estimates for the trace from equations \ref{fits0}. The case depicted is the same as the one shown in figure \ref{BetaFig}.  There are two adjustable constants in the fit.  First a constant, with the value 0.345 is subtracted from the numerically determined trace, so as to give an estimate of the deviation of the trace from its value at zero. Next, the asymptotic curves are both multiplied by a constant with value 1.05 so as to make the scale of the numerical trace the same as that of the asymptotic one.        }
\label{fit}
\end{figure}

\subsection{Check against numerical results}
To see how well this theory works, we have calculated the orbits, and from these the trace for the case in which $\xi(t)=8*(1-t)^\beta$ and $t_0=0$.  The general shape of the trace was shown in Figure \ref{BetaFig} above.

To check the asymptotic analysis, figure \ref{fit} plots the real and imaginary parts of the trace against the time to the singularity.  The curves on this figure are the lowest order estimates of these singularities
\begin{subequations}
\beq
\Re \Gamma_0(s) = \xi(s) + ( \Lambda(s))^{-1}  \log ( \Lambda(s)^{2} |s|/2)
\label{fits0r}
\eeq
\beq
\Im \Gamma_0(s) =2 \pi i \Lambda(s)
\eeq
\label{fits0}
\end{subequations}
The data is expected to fit the theory toward the left hand side of this figure, and indeed it does so.

\begin{figure}[t]
\includegraphics[scale=0.9]{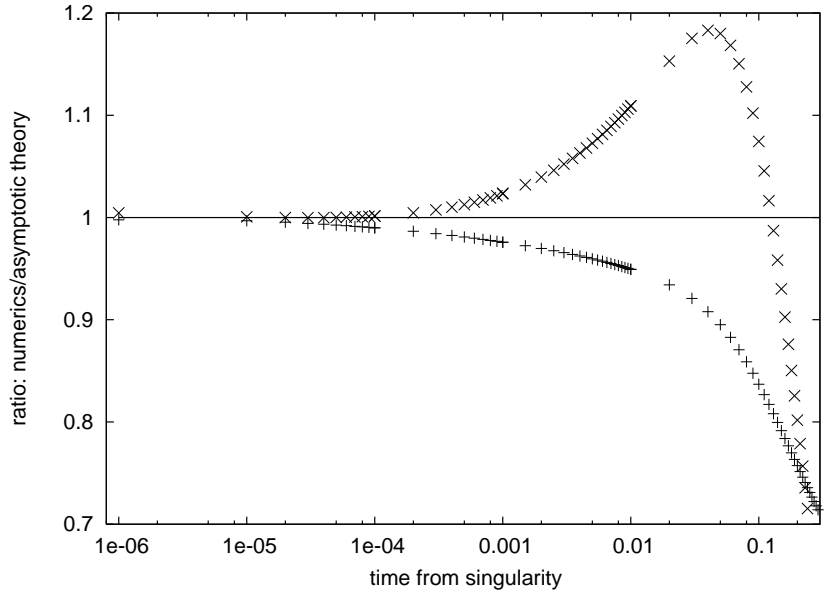}
%\vskip 1.5 in
%\includegraphics[scale=.35]{wplane.pdf}
%\vskip 1 in
\caption{Compensated traces.  This takes the same data as shown in Figure \ref{fit} and plots then in such a way as to emphasize the deviations from the theory. The real and imaginary parts of the calculated trace (minus the value at $s=0$) is compared with the theory as described in equation \ref{makefit}  The lower curve describes
$\Re(\gamma(s)-\gamma(0)) $ ; the upper is a plot involving $\Im(\gamma(s))$.  If theory is right, then the curves should asymptote to unity as time goes to zero. It does so in a satisfying fashion.  The fitting constants are $a=\gamma(0)=0.345, C=1.05$ and $d=0.05$.  }
\label{CompFig}
\end{figure}

However, log log plots like the one in figure \ref{fit} can often conceal difficulties. For example the second term in equation \ref{fits0r} is essentially invisible in figure \ref{fit}.  To make a more incisive comparison, take the numerically determined $\gamma(s)$ and argue that it can  be
fit by the asymptotic estimate $\Gamma_0$ of equation \ref{fits0} in the form
\beq
\gamma(s) -a=C  ~\Gamma_0(s) +d~ (\Gamma_0(s))^2
\label{makefit}
\eeq
Here $a, C$ and $d$ are real fitting constants.  The value of the trace at the singular point is given by $a$.   The magnification which might occur during the time well before the singularity is represented by the factor $C$.  A further non-linear magnification is given by $d$.  A sharp comparison is obtained by taking the real and imaginary parts of equation \ref{makefit} and diving the left hand sides by the right hand sides.  The result is shown in figure  \ref{CompFig}. An agreement between theory and numerics would be signaled by a set of points which approached unity as the time approached the value at the singularity.  The curves shown seem to support the asymptotic theory.

\section*{Acknowledgments}
We would like to thank Isabelle Claus for doing preliminary versions of some of the calculations reported here. We have had helpful conversations or correspondence with W. Kager, S. Rohde, Ilya Gruzberg,  This research was supported by the National Science Foundation's Division of Materials Research, the University of Chicago Materials Lab (MRSEC), and the ASCI/FLASH program of the department of energy.

\section*{Appendix: Computational Methods}
Two numerical schemes are employed in this paper.  One is the standard second order Runge-Kutta method \cite{NR} which we abbreviate as RK2.   The other method is a second order accurate version of the well-known \cite{MR} square root method, abbreviated, which uses
\begin{subequations}
\beq
(g_{t_f}(z)-\bar{\xi})^2= (g_{t_i}(z)-\bar{\xi})^2  +4(t_f-t_i) +O (t_f-t_i)^3
\label{SRM}
\eeq
The parameter, $\bar{\xi} $, is the average forcing, corrected to gain second order accuracy
\beq
\bar{\xi}= \xi((t_f +t_i)/2)  +\frac{t_f-t_i}{8} \big[   \frac{d\xi}{dt}(t_f)- \frac{d\xi}{dt}(t_i)  \big]
\eeq
\end{subequations}
To derive equation \ref{SRM} start from the exact relation:
$$
\frac{d}{dt} [ (g-\xi)^2-\xi^2 ] = 4-2\dot{\xi} g
$$
Integrate this between $t_1$ and $t_2$, using the trapezoidal approximation to estimate the right hand side.  One finds:
$$
 \Big[g^2-2\xi g -4t\Big]_1^2  = -(g_1~\dot{ \xi_1} +g_2~ \dot{\xi_2})(t_1-t_2)
$$
One then expands the $\xi$'s about the average time to obtain the result in equation \ref{SRM}

The step size is set by a parameter. STEP, with the actual step size being
\beq
(t_f-t_i)=\mbox{STEP} |\xi-g| / (d \xi /dt )
\eeq
with some cutoffs to prevent the actual difference, $(t_f-t_i)$, from getting too large or too small.

\end{document}